\documentclass[12pt]{article}
\usepackage{amsmath,amssymb,graphicx,cite,color,url,hyperref,setspace}
\def\beq{\begin{eqnarray}}
\def\enq{\end{eqnarray}}

\newcommand{\gsim}{\raisebox{-0.13cm}{~\shortstack{$>$ \\[-0.07cm] $\sim$}}~}
\def\lapp{\mathrel{\rlap{\raise.5ex\hbox{$<$}}{\lower.5ex\hbox{$\sim$}}}}
\def\gapp{\mathrel{\rlap{\raise.5ex\hbox{$>$}}{\lower.5ex\hbox{$\sim$}}}}
\begin{document}
\begin{flushright}
TIFR/TH/17-07
\end{flushright}

\vspace*{-0.4in}
\begin{center}
{\large{\sc Mixed Higgs-Radion States at the LHC -- a Detailed Study}} 

\setstretch{1.05}

Amit Chakraborty\footnote{E-mail: {\sf amit@post.kek.jp}} \\ [2mm]
{\small
Theory  Center,  Institute  of  Particle  and  Nuclear  Studies,  \\
KEK,  1-1  Oho,  Tsukuba,  Ibaraki 305-0801, Japan}

Ushoshi Maitra\footnote{E-mail: {\sf ushoshi@theory.tifr.res.in}}, 
Sreerup Raychaudhuri\footnote{E-mail: {\sf sreerup@theory.tifr.res.in}} 
{\small and} 
Tousik Samui\footnote{E-mail: {\sf tousik@theory.tifr.res.in}} \\ [2mm]
{\small
Department of Theoretical Physics, Tata Institute of Fundamental Research,  \\ 
1, Homi Bhabha Road, Mumbai 400\,005, India.}
\setstretch{1.15}

\medskip

\centerline{\today}

{\sc Abstract}
\end{center}
\vspace*{-0.35in}
\begin{quotation}
\small
\setstretch{1.05}
{\noindent Light radions constitute one of the few surviving possibilities for
observable new particle states at the sub-TeV level which arise in 
models with extra spacetime dimensions. It is already known that the 
125~GeV state discovered at CERN is unlikely to be a pure radion state, since its decays resemble those of the Standard Model Higgs boson too closely. However, due to experimental errors in the measured decay 
widths, the possibility still remains that it could be a mixture of 
the radion with one (or more) Higgs states. We use the existing LHC 
data at 8 and 13 TeV to make a thorough investigation of this possibility. Not surprisingly, it turns out that this model is already constrained quite
effectively by direct LHC searches for an additional scalar heavier than 125~GeV. 
We then make
a detailed study of the so-called `conformal point', where this
heavy state practically decouples from (most of) the Standard Model 
fields. Some projections for the future are also included.}
\end{quotation}
\centerline{PACS Nos:  04.60.Bc, 12.60.Fr, 14.80.Cp, 13.85.Rm}
\normalsize
\setstretch{1.25}

\noindent{\large\sc 1. Introduction}

The 2012 discovery\cite{Higgs}, at the LHC, of a weakly-interacting light scalar state 
--- which appears from all current indications to be an elementary Higgs particle --- revives the old question of how the mass of such a scalar can remain stable against large electroweak corrections in a theory with a momentum cutoff at some very high scale. This, as is well-known, goes by 
the name of the gauge hierarchy problem, or, alternatively, as the fine-tuning problem. It has also been known for several decades that any solution to this problem must invoke new physics beyond the Standard Model (SM) of strong and electroweak interactions. 

One of the most elegant solutions of the hierarchy problem is that devised 
in 1999 by L.~Randall and R.~Sundrum (RS)\cite{warped}. They considered a world with one 
extra space dimension, having the topology of a circle folded about a 
diameter ($\mathbb{S}^1/\mathbb{Z}_2$), at either end of which lies 
a pair of four-dimensional manifolds -- called `branes' -- containing 
matter. One of these is the
so-called infra-red (IR) brane, where all the SM fields lie, and the other 
is the so-called ultra-violet (UV) brane, where we have field elements 
comprising a theory of strong\footnote{Here `strong' means comparable to 
electroweak strength.} gravity. 
One can then tune the cosmological constant on the 
two branes, as well as that in the $\mathbb{S}^1/\mathbb{Z}_2$ bulk, to 
obtain a solution of the five-dimensional Einstein equations in the form 
of a `warped' metric
\begin{equation}
ds^2 = e^{-2{\cal K} R_c \phi} \eta_{\mu\nu} dx^\mu dx^\nu - R_c^2 d\phi^2
\label{eqn:RSmetric}
\end{equation}
where the $\mathbb{S}^1/\mathbb{Z}_2$ `throat' is characterised by the 
compactification radius $R_c$, an angular coordinate $\phi$ and a curvature parameter
${\cal K}$. It can then be shown that the mass of the Higgs scalar is 
generated on the UV brane at a value close to the bulk Planck mass $M_5$
(itself a little smaller than the four-dimensional Planck mass $M_P = 
\left(\hbar c/G_N\right)^{1/2}$), and projected on the IR brane through the 
expanding `throat', thereby acquiring the much smaller value 
\begin{equation}
M_H \sim e^{-\pi {\cal K} R_c} M_5
\label{eqn:RSHiggs}
\end{equation}   
If we can now tune ${\cal K} R_c \simeq 11.6$, we recover the correct ballpark
for the mass of the discovered scalar. This constitutes a neat solution to the hierarchy problem in terms of spacetime geometry, without having recourse to any parameters which are unnaturally large or small. In fact, the Planck scale
is the only fundamental mass scale in this theory.   

It is fair to ask, however, whether the parameter ${\cal K} R_c$ is protected 
against small dynamical fluctuations, for
\begin{equation}
\frac{\delta M_H}{M_H} \approx 11.6\pi \, \frac{\delta R_c}{R_c}
\label{eqn:RSfluc}
\end{equation}
i.e. small fluctuations in the inter-brane distance would lead to magnified
fluctuations in the Higgs boson mass. As the latter is now known to an accuracy 
of about 2\%, it follows that the inter-brane distance must be stable to an 
accuracy of about $5\times 10^{-4}$ --- for which the minimal RS model  
has {\it no} provision.

A brilliant solution to this was devised by Goldberger and Wise 
(1999)\cite{GWmechanism}. If one allows for fluctuations in the size of the extra dimension, 
we can rewrite the metric in Eq.~(\ref{eqn:RSmetric}) as
\begin{equation}
ds^2 = e^{- 2T(x) \phi} \eta_{\mu\nu} dx^\mu dx^\nu 
- \left[ \frac{T(x)}{\cal K} \right]^2 d\phi^2
\label{eqn:RSmodulus}
\end{equation}
where the dynamic $T(x)$ replacing ${\cal K}R_c$ is known as a modulus field. In the minimal RS model, this
is a free field and hence, as mentioned above, there is no constraint at all 
on ${\cal K} R_c = \langle T(x) \rangle$. Goldberger and Wise then augmented 
the model by the introduction of a bulk scalar $B(x,y)$, with a mass $M_B$ 
and quartic self-interactions on the IR and UV branes, with vacuum expectation values $V_{IR}$ and $V_{UV}$ respectively -- all 
these mass-dimension quantities being in the ballpark of the Planck mass. They were then able 
to show that the scalar modulus field $T(x)$ develops a potential 
with a minimum at
\begin{equation}
\langle T(x) \rangle = {\cal K} R_c \simeq \frac{4}{\pi} 
\left(\frac{{\cal K}}{M_B}\right)^2 \ln \frac{V_{UV}}{V_{IR}}
\label{eqn:GWminimum}
\end{equation}
which can be easily tuned to the required value $11.6$ by varying the unknowns
$M_B$, $V_{IR}$ and $V_{UV}$ without having recourse to unnaturally large 
or small numbers. This is consistent with the general philosophy of the RS model.

The modulus field $T(x)$, which is like a dilaton in the fifth dimension, can 
be parametrised as a {\it radion}
\begin{equation}
\varphi(x) = \Lambda_\varphi ~ e^{-\pi\left\{T(x) - {\cal K}R_c\right\}}
\label{eqn:Rdefinition}
\end{equation}
which has a vacuum
expectation value
\begin{equation}
\Lambda_\varphi = \sqrt{\frac{24M_5^3}{{\cal K}}}~e^{-\pi{\cal K}R_c}
\label{eqn:Rvev}
\end{equation}
and a mass
\begin{equation}
M^2_\varphi = \frac{2{\cal K}^2}{M_5^3} \left(V_{UV} - V_{IR}\right)^2 
e^{-2\pi{\cal K}R_c}
\label{eqn:Rmass}
\end{equation}
Because of the warp factor $e^{-\pi{\cal K}R_c}$ , both the radion mass $M_\varphi$ 
and the radion vacuum
expectation value $\Lambda_\varphi$ lie at or around the electroweak scale. Hence, 
it is easier, for phenomenological purposes, to treat them as the free parameters 
in the theory, rather than the set $\left\{{\cal K}, M_5, V_{UV}, V_{IR}\right\}$. 
It is also worth noting that if we let $V_{UV} = V_{IR}$, in which 
case Eq.~(\ref{eqn:Rmass}) tells us that the radion is massless, we would also 
have $R_c = 0$ from Eq.~(\ref{eqn:GWminimum}), i.e. the two branes would coalesce
and $M_H$ immediately shoot up to $M_5$ --- which takes us back to the Standard Model 
and the hierarchy problem. We conclude, therefore, that $V_{UV} > V_{IR}$ and hence 
the radion must be massive. 

The interactions of the radion with matter on the IR brane will naturally 
follow those of the dilaton (which it is a variant of) and can be written as
\begin{equation}
{\cal L}_{\rm int}(\varphi) = \frac{1}{\Lambda_\varphi} \, \varphi \left( T_\mu^\mu 
+ {\cal A}_T \right)
\label{eqn:Rint}
\end{equation} 
where $T_{\mu\nu}$ is the tree-level energy-momentum tensor and ${\cal A}_T$ 
is the trace anomaly. For on-shell particles, the tree-level $T_\mu^\mu $ has 
the explicit form 
\begin{equation}
T_\mu^\mu  = \sum_f m_f \bar{f} f + M_H^2 H^2 - 2M_W^2 W^{+\mu}W^-_\mu 
- M_Z^2 Z^\mu Z_\mu   
\label{eqn:trEMtensor}
\end{equation}
where the sum runs over all fermions $f$. This, apart from the 
${\cal A}_T$ term,  is exactly like the coupling of 
the Higgs boson, except that the SM vacuum expectation value $v$ is replaced 
by the radion vacuum expectation value $\Lambda_\varphi$. Not surprisingly, 
radion phenomenology is very similar to Higgs boson phenomenology. It differs,
however, in the anomaly term
\begin{equation}
{\cal A}_T = \sum_i \frac{\beta(g_i)}{2g_i} \, F^{\mu\nu i} F_{\mu\nu}^i
\label{eqn:trAnomaly}
\end{equation}
where $\beta(g_i)$ is the beta function corresponding to the coupling $g_i$
of the gauge field $A_i$ which has the field strength tensor $F_{\mu\nu}^i$.
The sum over $i$ runs over all the gauge fields in the SM, including photons,
gluons and $W^\pm$ and $Z$ bosons. The ${\cal A}_T$ term induces 
substantial couplings of the 
radion to $\gamma\gamma$ and $gg$ pairs, which are completely absent in 
Eq.~(\ref{eqn:trEMtensor}). On the other hand, similar anomaly-induced 
contributions to radion couplings with $W^+W^-$ and $ZZ$ pairs are 
usually negligible 
compared to the corresponding terms in Eq.~(\ref{eqn:trEMtensor}), because
of the large masses of these particles, and only become significant when their tree-level couplings to one of the scalars vanishes.  

Like the Higgs boson, the tree-level radion couplings in Eqn.~(\ref{eqn:Rint}) would
be subject, in addition to the trace anomaly contributions, to radiative corrections,
especially from loops involving the top quark. Moreover, it is worth mentioning that there could be large brane corrections to the above couplings if the mass of the radion 
is comparable to the Kaluza-Klein scale \cite{Chacko}, determined by the mass of the 
lightest graviton mode in the minimal RS construction. To avoid this, we require a radion 
which is comparatively light, and this requires a modest level of fine tuning 
\cite{Chacko}. The discussions in this article are, therefore, subject to this assumption.

As remarked above, the phenomenological behaviour of such a light radion is rather
similar to that of the Higgs boson. This naturally leads one to ask whether
these two low-lying elementary scalar states can {\it mix}, since they carry the 
same set of conserved quantum numbers, once the electroweak symmetry has been broken. In fact, this is possible, as was first pointed out in
Ref.~\cite{R-H-mixing} and has been discussed by many others\cite{R-H-mixing,R-H-mixing1,R-H-mixing2}. 
Before proceeding further, it may be noted that there are several  phenomenological models with fermions and gauge bosons accessing the bulk\cite{Huber:2000ie, Gherghetta:2000qt, Grossman:1999ra, Agashe:2006at, Iyer:2015ywa}, which have better control over the flavour problem. In these models, the top quark remains close to the TeV brane along with the Higgs field while the other fermions are close to the UV brane. This suppresses the 
higher-dimensional operators contributing to flavour-changing neutral currents, since the effective interaction of fermions with the Higgs field is governed by the overlap of their profiles and hence this scenario 
naturally generates the 
pattern of fermion masses and mixings. These models predict heavy Kaluza-Klein particles on the TeV brane having masses in the range of a TeV. However, the radion and Higgs fields, being still close to the TeV brane, 
mix more-or-less without bulk effects~\cite{bulk_radion}. 
Hence, the mixing can be understood fairly 
accurately using a minimal model 
where all the relevant particles are confined to the TeV brane\footnote{The only caveat to this is the fact that heavy Kaluza-Klein excitations of the
top quark may contribute to Higgs production at a hadron collider through loop diagrams. However, if these excitations are at the level of a TeV, the corresponding loop
contributions are not more than a few percent and may be safely neglected
--- as we have done in this work.}, for this is, after all, no more than
approximating a sharply-peaked function by a delta function. 

In the following section, therefore, we briefly discuss, following Refs.~\cite{R-H-mixing1,R-H-mixing2} how the radion-Higgs field mixing may be 
described in terms of a single
mixing parameter $\xi$. The next section then describes constraints on the
mixed Higgs-Radion scenario, as obtained using all experimental inputs
currently available, especially those from the LHC. For easy
comparison, we include projections of the discovery reach of the LHC alongside
the current constraints. Before concluding, we include a short section on
the so-called `conformal point' near $\xi = 1/6$, which has unique features. 
While some of the observations in this paper echo previous ones\cite{R-H-pheno}, the data
used are current, leading to new bounds, and, for ease of reading, we have
presented our findings in a manner such that this paper can be read as far
as possible independently of the preceding literature.  

\bigskip

\noindent{\large\sc 2. Radion-Higgs mixing}

Mixing of the radion field $\varphi(x)$ with the Higgs scalar $h(x)$ of the SM
has been discussed by several authors~\cite{R-H-mixing,R-H-mixing1,R-H-mixing2}, with the same broad
features, but we choose to closely follow the formalism of 
Ref.~\cite{R-H-mixing1,R-H-mixing2}.

The mixing occurs through the kinetic terms
\begin{equation}
{\cal L} = \frac{1}{2} \partial^\mu h \, \partial_\mu h 
- \frac{1}{2} M_{h}^2 h^2                                                                                                                                                                                                                                                                                                        + \frac{\beta}{2} \partial^\mu \varphi \,
\partial_\mu \varphi - \frac{1}{2} M_\varphi^2 \varphi^2                                                                                                                                                                                                                                                                                                        
 + 6\gamma\xi \, \partial^\mu \varphi \, \partial_\mu h  
 \label{eqn:Lagrangian}
\end{equation}
where $\gamma \equiv v/\Lambda_\varphi$, $v$ being the SM Higgs vacuum
expectation value. In this
formalism, the mixing parameter appears twice -- once in the mixing term
$6\gamma\xi \, \partial^\mu \varphi \, \partial_\mu h$, and once in the
non-canonical normalisation $\beta \equiv 1 + 6\gamma^2\xi$ of the radion 
kinetic term. As is usual, the Higgs boson mass is given by 
$M_h^2 = 2\lambda v^2$, 
where $\lambda$ is the Higgs quartic coupling and $v$ is the Higgs vacuum
expectation value. 

We note that the presence of the non-canonical normalisation $\beta$ means that
the identification of physical states $H$ and $\Phi$ will involve a scaling
as well as a rotation of states, i.e. a non-unitary transformation. Hence, we 
write the unphysical states $\varphi, h$ as linear combinations of the
physical ones $\Phi, H$, with real coefficients $A,B,C$ and $D$, thus
\begin{eqnarray}
\varphi & = & A\,\Phi + B\,H \nonumber \\
h & = & C\, \Phi + D\,H \ ,
\label{eqn:mixing}
\end{eqnarray}
where the coefficients $A,B,C$ and $D$ are given by
\begin{eqnarray}
A & = & -\frac{1}{Z} \cos \theta \qquad\qquad\qquad\quad
B   =  \frac{1}{Z} \sin \theta \nonumber \\
C & = & \sin \theta + \frac{6\gamma\xi}{Z} \cos \theta \qquad\qquad
D   =   \cos \theta - \frac{6\gamma\xi}{Z} \sin \theta
\label{eqn:coefficients}
\end{eqnarray}
in terms of
\begin{equation}
Z^2 = \beta - (6\gamma\xi)^2
\label{eqn:Zparam}
\end{equation}
and a mixing angle $\theta$, defined by
\begin{equation}
\tan 2\theta = \frac{12\gamma\xi Z M_h^2}
{M_\varphi^2 - M_h^2 \left(Z^2 - 36\gamma^2\xi^2\right)}
\label{eqn:mixangle}
\end{equation}
The mixing parameter $\xi$ is immediately constrained by the requirement that
$Z^2 > 0$ to get a real mixing angle.
The mass eigenvalues of the physical eigenstates $\Phi$ and $H$ 
are now given by
\begin{equation}
M^2_{\Phi,H} = \frac{1}{2Z^2} \left( M^2_\varphi + \beta M^2_h
\pm \sqrt{\left(M^2_\varphi + \beta M^2_h\right)^2 - 4Z^2 M^2_\varphi M^2_h} 
\right) 
\label{eqn:scalarmasses}
\end{equation}
where the sign is chosen to ensure that $M_H < M_\Phi$. We identify the
lighter state $H$ as the scalar state of mass around 125~GeV which was 
discovered at the CERN LHC in 2012, while the other state $\Phi$ is
a heavier scalar state predicted in the model.
From these formulae, it is clear that the free parameters in question are
$M_h$, $M_\varphi$, $\Lambda_\varphi$ and $\xi$, everything else being 
computable in terms of them. We also note in passing that since $M^2_h = 
2\lambda v^2$, this makes 
the Higgs quartic coupling $\lambda$ an unknown quantity in this model, just 
as it used to 
be in the Standard Model before the identification of the 125~GeV scalar
with the Higgs boson\footnote{This is a reflection of the fact that we
still do not have a direct measurement of $\lambda$. All that we have is the 
estimate $\lambda = (125~{\rm GeV})^2/2v^2 \simeq 0.129$ --- which is true only if the 125~GeV state is purely a SM Higgs boson without any admixture of new states.}. 

Instead of the Lagrangian parameters $M_h$ and $M_\varphi$, however, we 
find it 
more convenient to use the physical masses $M_H$ and $M_\Phi$, which can 
be traded for the previous two by some simple algebra, leading to
\begin{eqnarray}
M_\varphi^2 = \frac{Z^2}{2}\left[ M_\Phi^2 + M_H^2 +
\sqrt{(M_\Phi^2 + M_H^2)^2 - \frac{4\beta M_\Phi^2 M_H^2}{Z^2}} \right]
\nonumber \\ 
M_h^2 = \frac{Z^2}{2\beta}\left[ M_\Phi^2 + M_H^2 -
\sqrt{(M_\Phi^2 + M_H^2)^2 - \frac{4\beta M_\Phi^2 M_H^2}{Z^2}} \right]
\label{eqn:unscalarmasses}
\end{eqnarray}
Since we 
identify $M_H = 125$~GeV, we are left with a set of only three independent 
parameters, viz. $M_\Phi$, $\Lambda_\varphi$ and $\xi$. The rest of our 
analysis will be presented in terms of these variables.

We now have another theoretical constraint, apart from $Z^2 > 0$. This
is the requirement that the parameters $M_\varphi$ and $M_h$ be real 
(to keep the Lagrangian Hermitian), which automatically means that
\begin{equation}
\left(M_\Phi^2 + M_H^2\right)^2 > \frac{4\beta M_\Phi^2 M_H^2}{Z^2}
\label{eqn:discriminant}
\end{equation}
Imposing both these constraints reduces the possible range of $\xi$,
for a given $M_\Phi$ and $\Lambda_\varphi$, quite significantly 
(see below).

\begin{center}
\begin{figure}[!htb]
\centerline{\includegraphics[width=\textwidth]{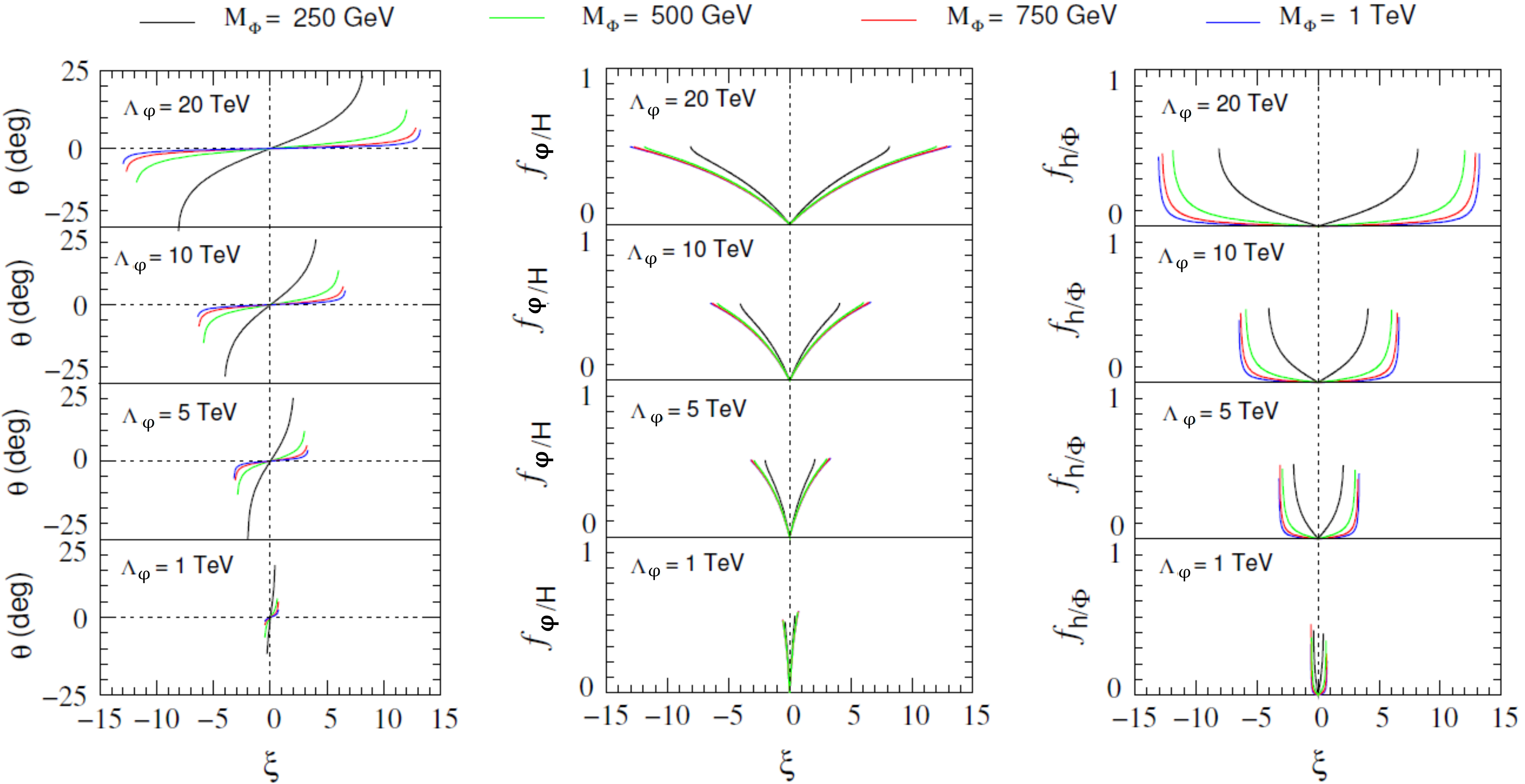} }
\vspace*{-0.1in}
\caption{\setstretch{1.05}\footnotesize The variation with $\xi$ of the mixing parameters ($a$) $\theta$, 
($b$) $f_{\varphi/H}$ and ($c$) $f_{h/\Phi}$. In each panel, the four boxes, from 
bottom to top, show the behaviour when $\Lambda_\varphi = 1$, 5, 10 and 20~TeV 
respectively, as marked. Inside the boxes, the curves are coloured black, green, 
red and blue for $M_\Phi = 250$~GeV, 500~GeV, 750~GeV and 1~TeV respectively. 
Observe that all these parameters vanish when $\xi = 0$, as expected. The lines 
break off abruptly for larger values of $|\xi|$ because of the theoretical 
constraints discussed in the text.}
\label{fig:mixing}
\end{figure}
\end{center}

Since the mixing of the $h$ and the $\varphi$ to produce the physical $H$ and
the $\Phi$ is non-unitary, we define two mixing indicators as follows.
We first invert Eq.~(\ref{eqn:mixing}) to write
\begin{eqnarray}
\Phi & = & a\,\varphi + b\,h \nonumber \\
H & = & c\, \varphi + d\,h \ ,
\label{eqn:mixing}
\end{eqnarray}
where
\begin{equation}
\left( \begin{array}{cc} a & b \\ c & d \end{array} \right) 
= \left( \begin{array}{cc} A & B \\ C & D \end{array} \right)^{-1} \ . 
\end{equation}
In terms of this, we now define indicators
\begin{equation}
f_{\varphi/H} = \frac{|c|}{|c|+|d|}
\qquad\qquad
f_{h/\Phi} = \frac{|b|}{|a|+|b|}
\label{eqn:ratios}
\end{equation}
which, in a sense, indicate the fraction of radion $\varphi$ in the light state $H$, and the fraction of Higgs boson $h$ in the heavy state $\Phi$. These, together with the mixing angle $\theta$ defined in Eq.~(\ref{eqn:mixangle}), are plotted in Fig.~\ref{fig:mixing}, as a function of the mixing parameter $\xi$. 
\vspace*{-0.5in}

In each of the three panels in Fig.~\ref{fig:mixing}, we have four boxes placed
one above the other, corresponding to choices of four
different values of the radion vacuum
expectation value, viz. $\Lambda_\varphi = 1, 5, 10$ and 20 TeV respectively (marked in the respective boxes). Within each box, the curves
are colour-coded, with black, green, red and blue indicating benchmark 
choices of the
heavy scalar mass as $M_\Phi = 250$~GeV, $500$~GeV, $750$~GeV and 1~TeV
respectively (indicated at the top of the figure). Each curve ends abruptly
at some maximum and minimum values of the mixing parameter $\xi$ -- this is 
a reflection of the theoretical limitations (see above). As may be seen from the different plots, this restriction
is extremely stringent when $\Lambda_\varphi$ is small, and even when we push 
$\Lambda_\varphi$  as high as 20, does not permit the value of $|\xi|$ to exceed
15. If we consider the panel on the left, it is clear that we get significant 
values of the mixing angle $\theta$ only when the heavy $\Phi$ state is as light 
as around 250~GeV. For values of $M_\Phi$ of 500~GeV or greater, $\theta$ does 
not exceed $10^0$. However, since the mixing is not unitary, the smallness of 
$\theta$ is not necessarily an indicator of small mixing. This becomes clear
if we look at the central and right panels of Fig.~\ref{fig:mixing}, which
tell us the proportion of the radion in the 125~GeV state, and the proportion of
the Higgs boson in the heavier state respectively. In each case, as $|\xi|$
increases, the mixing becomes more, starting from zero when $|\xi| = 0$ to
about equal mixtures when $|\xi|$ reaches its maximum theoretically-permitted value. The purpose of this paper is, as explained above, to 
see how far such large mixings are allowed in the light of current experimental data. 
  
We next consider the effect of mixing on the couplings of the two scalar
states to the SM fields. As shown in Ref.~\cite{R-H-mixing2}, the tree-level
couplings 
of the heavy $\Phi$ state to pairs of SM fields $X\bar{X}$ (except
$X = H$) have the form
\begin{equation}
g_{\Phi X\bar{X}} = g_{\varphi X\bar{X}} \left( C + \gamma A \right)
\equiv c_\Phi \ g_{\varphi X\bar{X}} 
\label{eqn:cR}
\end{equation}
where $g_{\varphi X\bar{X}}$ can be read off from Eqs.~(\ref{eqn:Rint} --\ref{eqn:trEMtensor}), and $c_\Phi = C + \gamma A$ is a scaling factor. 
Similarly, the couplings of the light 125~GeV state have the form
\begin{equation}
g_{H X\bar{X}} = g_{h X\bar{X}} \left( D + \gamma B \right)
\equiv c_H \ g_{h X\bar{X}} 
\label{eqn:cH}
\end{equation} 
where $g_{h X\bar{X}}$ are the SM couplings and $c_H = D + \gamma B$ is
a scaling factor. Very different from these is the coupling of the heavy
scalar to a pair of light scalars, since all three fields 
are mixed states, and this
can be written~\cite{R-H-mixing2} for a $\Phi(p)-H(k_1)-H(k_2)$ vertex, as
\begin{eqnarray}
g_{HH} & = & \frac{1}{\Lambda_\varphi} \left[ \left(k_1^2 + k_2^2 \right)
\left\{ AD^2 + 6\xi B\left( CD + \gamma AD + \gamma BC \right) \right\} \right.  \\
&& + \left. D\left\{ 12\gamma\xi AB + 2BC + (6\xi - 1)AD  \right\} p^2
- 4M_h^2 D(AD + 2BC) - 3 M_h^2CD^2/\gamma \right] \nonumber
\end{eqnarray}
The couplings of the scalars $H$ and $\Phi$ with other particles are
conveniently listed in the Appendix of Ref.~\cite{R-H-mixing2}. 
 
\begin{figure}[!htb]
\centerline{\includegraphics[width=\textwidth]{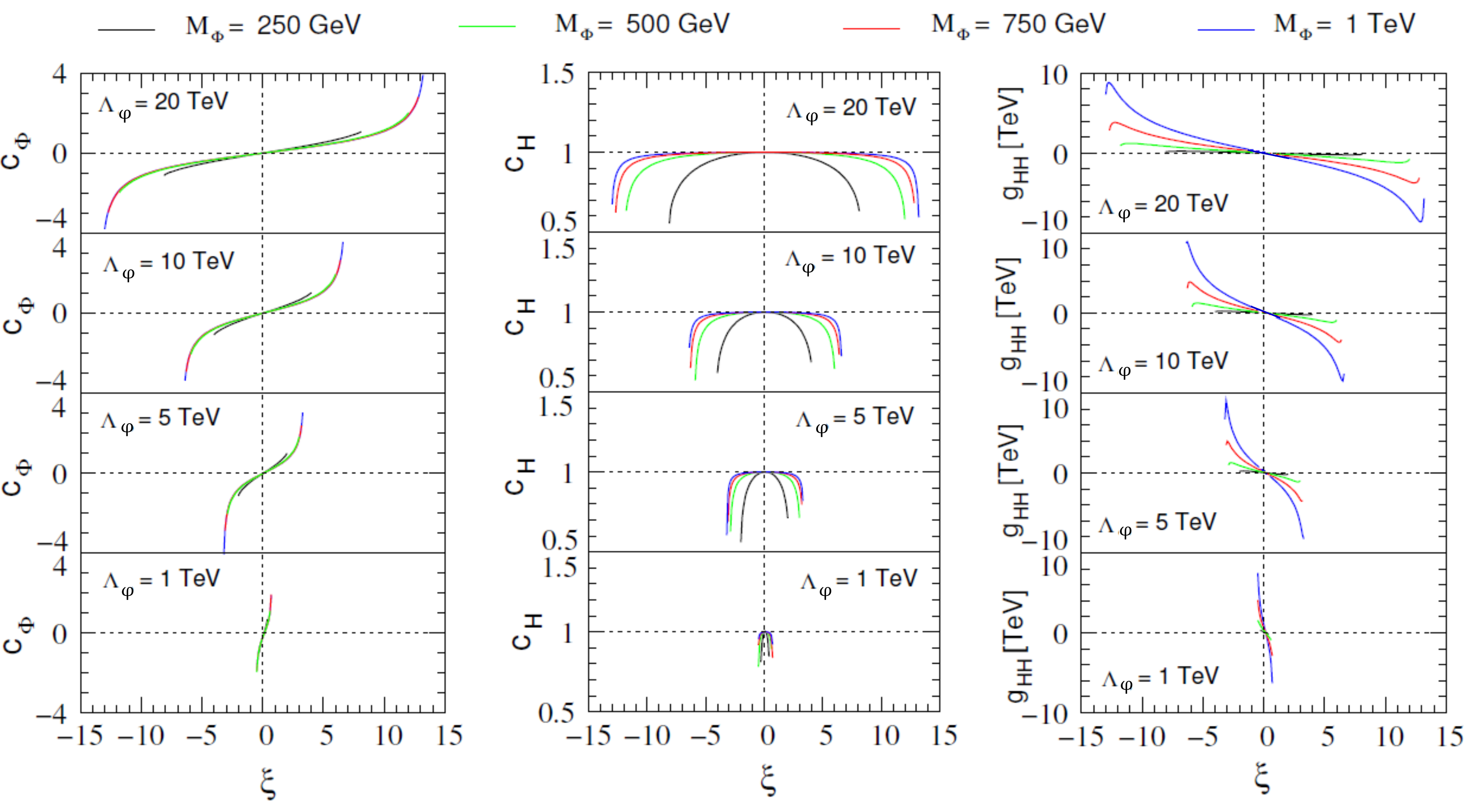} }
\vspace*{-0.1in}
\caption{\setstretch{1.05}\footnotesize The variation with $\xi$ of the (dimensionless) scaling factors ($a$) $c_\Phi$
and ($b$) $c_H$ is shown in the left and central panels, while the
right panel shows the $\Phi HH$ coupling $g_{HH}$, in units of TeV. 
The layout and colour conventions of this figure closely follow those 
of Fig.~\ref{fig:mixing}.} 
\label{fig:coupling}
\end{figure}

To get a feeling of how these couplings are affected by the variation
in the basic parameters $\xi$, $\Lambda_\varphi$ and $M_\Phi$, we plot
them in Fig.~\ref{fig:coupling} on a scheme similar to that in Fig.~\ref{fig:mixing}. The three panels show, from left to right, the 
scaling factors $c_\Phi$ and $c_H$, and the coupling $g_{HH}$ respectively.
As in Fig.~\ref{fig:mixing} it is immediately clear that for $\xi = 0$,
$c_\Phi$ is very small (small enough to appear as zero on this scale), 
as befits a radion with a small coupling to matter, whereas $c_H = 1$ indicating that the lighter scalar is the SM Higgs boson. Similarly,
for $\xi = 0$, 
the $g_{HH}$ coupling is very small (small enough to appear as zero on 
this scale), indicating that the heavy scalar couples only weakly to a 
pair of light scalars. There are also genuine zeroes in the couplings, which
are discussed in more detail in Section~4.     
 
An interesting feature of both Fig.~\ref{fig:mixing} and Fig.~\ref{fig:coupling} is the fact that the variation in parameters
is rather slow for smaller values of $\xi$, but is very sharp
for larger values just before the unphysical region. These larger
values of the scaling factor and $\Phi HH$ coupling are likely to
have phenomenological consequences at observable levels, and hence
are more likely to be constrained by experimental data. In the
next section, we shall see that this is indeed the case.

\bigskip

\noindent{\large\sc 3. Experimental Constraints}

We are now in a position to apply the experimental constraints to
this model. Since the two scalars $H$ and $\Phi$ are the crucial elements, 
the main constraints will come from
\vspace*{-0.2in}
\begin{enumerate}
\item[($a$)] the measured signal strengths $\mu_{XX}$ of the 125~GeV
scalar in its decay channels to $X\bar{X}$ pairs -- these are known to
match reasonably closely to the SM predictions, leaving only limited room
for a mixed state;
\item[($b$)] the lack of signals for a heavy scalar in the range of a few hundred
GeV to about a TeV -- by implication, any new scalar would be very heavy
and mix only marginally with the SM Higgs boson. 
\end{enumerate} 
\vspace*{-0.2in}
In principle, the scalars could also contribute as virtual states to 
any neutral current processes. However, as most of these are suppressed
by the small masses of the initial states (either $e^\pm$ or $u$ and $d$ 
quarks), we do not really get any useful constraints from these processes.
Constraints from electroweak precision tests are not very strong 
\cite{R-H-mixing1,precision}.
In the rest of this sections, therefore, we concentrate on the two
issues listed above.

\begin{table}[!h]
\begin{center}
\begin{tabular}{|c|c|c|}
\hline
Signal Strength &  8 TeV limits & 13 TeV limits \\
\hline\hline
& & \\ [-4mm]
$\mu_{\gamma\gamma}$ & 0.68 -- 1.70\cite{SS_ATLCMS_comb} & $\left\{\begin{array}{c} 
0.31 - 1.27\text{\cite{SS_CMS_gmgm_13tev}} \ ({\rm CMS})~~~~ \\ [1mm]
0.03 - 1.17\text{\cite{SS_ATL_gmgm_13tev} \ ({\rm ATLAS}}) \end{array} \right.$ \\ [1 mm]
\hline
& & \\ [-4mm]
$\mu_{WW}$           & 0.58 -- 1.42\cite{SS_ATLCMS_comb} &  --- \\ 
\hline
& & \\ [-4mm]
$\mu_{ZZ}$           & 0.76 -- 2.16\cite{SS_ATLCMS_comb} & 0.78 -- 1.62\cite{SS_CMS_ZZ_13tev} (CMS) \\ 
\hline
& & \\ [-4mm]
$\mu_{\tau\tau}$     & 0~~~    -- 2.26\cite{SS_ATLCMS_comb} & --- \\ 
\hline
& & \\ [-4mm]
$\mu_{bb}$           & 0~~~    -- 3.13\cite{SS_ATLCMS_comb} & 0 --- 1.23 ({\rm ATLAS})\\ 
\hline
\end{tabular}
\caption{\footnotesize LHC results on the Higgs signals strengths at 95\% confidence level.
The 8~TeV limits are from ATLAS and CMS combined. Production is through
gluon fusion, except for the last entry, which is through vector boson
fusion.} 
\label{tab:SS}
\end{center}
\end{table}

We first take up the signal strengths of the 125~GeV scalar $H$. This decays
into several channels
\begin{equation}
H \longrightarrow X + \bar{X}
\end{equation}
where $X = \ell^-, u, d, s, c, b, W, Z, \gamma, g$ with one of $X$ or 
$\bar{X}$
being off-shell in the case of $W$ and $Z$. At the LHC, the $H$ is produced 
dominantly through gluon-gluon fusion\footnote{In our numerical analysis, we 
have also included the vector boson fusion mode.}. Hence, we can define signal 
strengths $\mu_{XX}$ as
\begin{equation}
\mu_{XX} = 
\frac{\sigma(pp \to gg \to H)_{\rm exp} \ {\mathcal B}(H \to X\bar{X})_{\rm exp}}
     {\sigma(pp \to gg \to H)_{\rm SM} \ {\mathcal B}(H \to X\bar{X})_{\rm SM}}
\label{eqn:SS}
\end{equation}
where $\sigma$ and ${\cal B}$ stand for cross-section and branching 
ratio respectively, and the subscripts `SM' and `exp' mean the SM 
prediction and the experimental 
value respectively. If we are making a theoretical prediction, then `exp' will
stand for the expected value in the theoretical model in question --- in the 
present case, the model with radion-Higgs mixing. Of course, in an experiment
only the entire numerator on the right side of Eq.~(\ref{eqn:SS}) can be 
measured and not the individual factors. By this definition, then, all the
SM signal strengths are normalised to unity, and experimental deviations from 
it constitute the leeway for new physics. These allowed experimental deviations 
are given in Table~\ref{tab:SS}.

\begin{figure}[!h]
\centerline{\includegraphics[width=\textwidth]{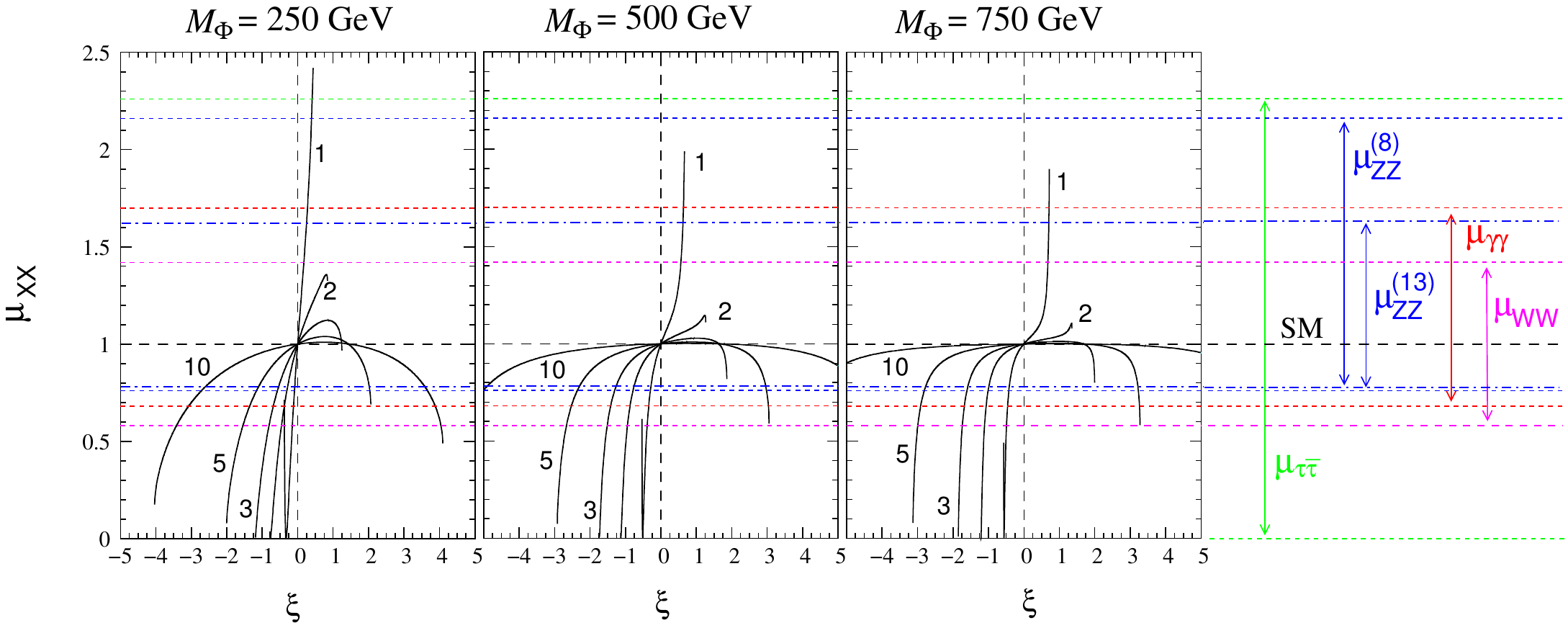} }
\vspace*{-0.1in}
\caption{\setstretch{1.05}\footnotesize The variation of the predicted signal strengths
with the mixing parameter $\xi$, for different choices of 
$\Lambda_\varphi$ (in TeV), marked alongside each curve. Each panel corresponds to a different mass $M_\Phi$ as
marked. The experimental constraints at 95\% C.L. are shown on the right.
Superscripts (8) and (13) indicate results from Run-1 and Run-2 respectively
of the LHC.}
\label{fig:signalstrengths}
\end{figure}

Obviously, for zero mixing, the signal strengths predicted for the $H$
scalar will be the same as the SM values, i.e. unity. As $\xi$ increases,
we  should expect deviations from unity, and indeed that is what happens,
as illustrated, in Fig.~\ref{fig:signalstrengths}. The three panels, from
left to right, correspond to choices of $M_\Phi =$ 250, 500 and 750 GeV 
respectively. The graph for $M_\Phi = 1~TeV$ is very similar to that for 
$M_\Phi = 750$ GeV, and hence we do not show it explicitly. Likewise,
the actual graphs for $\mu_{\gamma\gamma}$ are slightly different, but not 
enough to show up on a plot at this scale. Each curve
in the panels corresponds to the value of $\Lambda_\varphi$, in TeV, written
alongside, i.e. 1, 2, 3, 5 and 10 TeV respectively. The steepness of
the curves decreases with increasing $\Lambda_\varphi$, for which we also
have larger permitted ranges in $\xi$, as we have earlier shown in
Fig.~\ref{fig:coupling}. Horizontal broken lines in 
Fig.~\ref{fig:signalstrengths} represent the useful 95\% C.L. constraints from
the signal strengths in Table.~\ref{tab:SS}, and are marked on the 
right side of the figure.  

The behaviour of the predicted signal strengths with increasing $\xi$ is 
quite as expected, remaining close to the SM value for small $\xi$ and
showing large deviations near the edge of the theoretically-allowed range.
This, as we have seen earlier, is due to the large deviations of the coupling
of the $H$ from the SM coupling at such values of $\xi$. It is thus obvious
that the present constraints from signal strengths will only affect narrow
strips of the parameter space adjacent to the theoretically-disallowed region,
and this, in fact, is what we find (see below). It may be noted in passing
that a region of the parameter space where $D + \gamma B \simeq 0$ would be
very strongly constrained from the signal strengths, but this does not happen
anywhere inside the region allowed by theoretical considerations.  

\begin{figure}[!htb]
\centerline{\includegraphics[width=\textwidth]{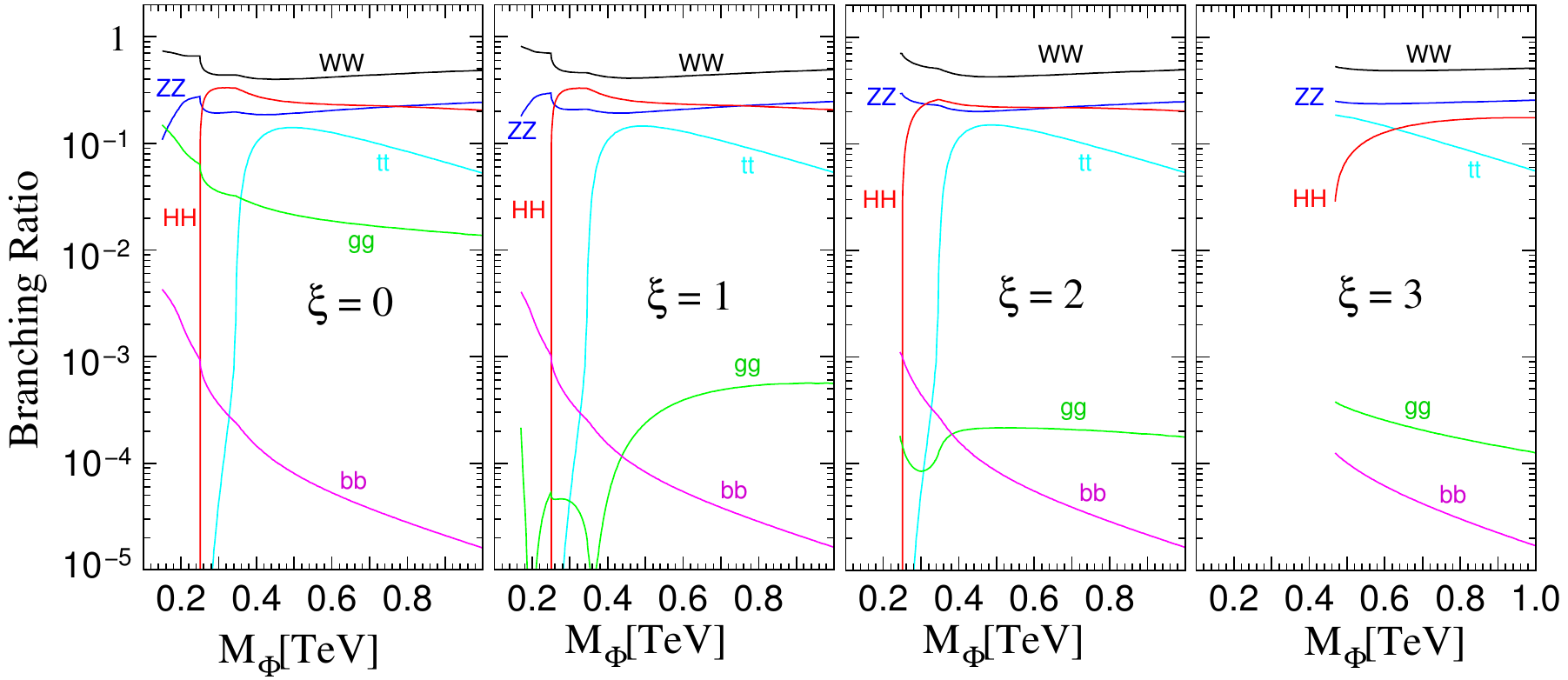} }
\caption{\setstretch{1.05}\footnotesize Two-body branching ratios of the 
heavy scalar
$\Phi$ as a function of its mass $M_\Phi$, for different choices of the
mixing parameter $\xi = 0$, 1, 2 and 3. The extreme left panel, viz. $\xi = 0$,
corresponds to a pure radion state. Branching ratios for the diphoton channel 
are not shown as they are too small to appear on the chosen scale. For these 
plots, we have set $\Lambda_\varphi=5$~TeV. Variation with 
$\Lambda_\varphi$ exists, but is slight. }
\label{fig:RadionBR}
\end{figure}

When we turn to the heavy $\Phi$ state, once again the main production mode 
is through gluon-gluon fusion, but now there is no analogous SM prediction and hence one looks for the direct signals in
the various decay channels of the $\Phi$. As in the case of the light scalar, the
potentially observable ones are $\Phi \to \gamma\gamma$\cite{HH_CMS_gmgm_8tev,HH_ATL_gmgm_13tev,HH_CMS_gmgm_13tev},
$WW$\cite{HH_CMS_WWandZZ_8tev,HH_ATL_WW_8tev,HH_ATL_WW_13tev,HH_CMS_WW_13tev},
$ZZ$ \cite{HH_ATL_ZZ_8tev,HH_CMS_WWandZZ_8tev,HH_ATL_ZZ_13tev,HH_CMS_ZZ_13tev} and 
$\tau^+\tau^-$\cite{HH_ATL_tautau_8tev,HH_CMS_tautau_8tev,HH_ATL_tautau_13tev,HH_CMS_tautau_13tev} to which we can now add $\Phi \to t\bar{t}$ and $\Phi \to HH$\cite{HH_ATL_hh_8tev,HH_CMS_hh_8tev,HH_ATL_hh_13tev,HH_CMS_hh_13tev}.
The $b\bar{b}$\cite{HH_CMS_bb} signal would be difficult to distinguish from the QCD
background, unless the mass of the $\Phi$ scalar is very well known, as in the 
case of the $H$ scalar. The behaviour of all these branching ratios, as
functions of the scalar mass $M_\Phi$ is shown in Fig.~\ref{fig:RadionBR},
where $\Lambda_\varphi$ is fixed to 5 TeV and the panels, from left to right,
correspond to $\xi = 0$ (no mixing), and $\xi = 1$, 2 and 3 respectively.
The relevant decay channel is marked alongside each curve. These curves
terminate at the left end where they correspond to theoretically-disallowed 
regions in the parameter space. 

One feature which is immediately obvious from these curves is the fact
that the scalar $\Phi$ decays dominantly through the $WW$ and $ZZ$ channels.
When the mixing is low, the $HH$ channel is also competitive, but as $\xi$
rises, it gets suppressed. In any case, the signals from the $WW$ and $ZZ$
channels are leptonic and clean, whereas the signals arising from $HH$,
dominantly leading to $4b$ final states, are hadronic, as are those arising
from the direct decays of 
the $\Phi$ into quark pairs. These hadronic channels are generally suppressed 
compared to $WW$ and $ZZ$, and, in any case, would be plagued by large QCD 
backgrounds. It may be still possible to investigate the $t\bar{t}$ and $HH$
channels, using jet substructure-based tagging methods for boosted particles, 
but such experimental searches are still not competitive 
\cite{ATLAS:2016ixk}. Thus, in principle, we get 
constraints from every decay channel of the $\Phi$, but the most
useful ones will arise from  the ATLAS and CMS search results for a heavy 
scalar resonance decaying
to $WW$ and $ZZ$ pairs, which are equally applicable to the $\Phi$ scalar
in the model under consideration. As is well-known, the experimental results are all negative, and hence the 95\% C.L. upper limits on the cross-section 
are given in Table~\ref{tab:HH}.
\begin{table}[!h]
\begin{center}
\begin{tabular}{|c|c|c|c|c|c}
\hline
$pp \to S \to WW$ & $M_S = 250$ GeV & $M_S = 500$ GeV & $M_S = 750$ GeV & $M_S = 1$ TeV  \\
\hline\hline
& & & & \\ [-3mm]
ATLAS (Run I)\cite{HH_ATL_WW_8tev} & --- & 0.191  & 0.039 & 0.020  \\ [2mm]
\hline  
& & & & \\ [-3mm]
CMS (Run I)\cite{HH_CMS_WWandZZ_8tev} & 1.590  & 0.287 & 0.221 & 0.064 \\ [2mm]
\hline 
& & & & \\ [-3mm]
ATLAS (Run II)\cite{HH_ATL_WW_13tev} & ---  & 0.884  & 0.253 & 0.066  \\ [2mm]
\hline 
& & & & \\ [-3mm]
CMS (Run II)\cite{HH_CMS_WW_13tev} & 51.395 & 4.866  & 2.882 & 1.708  \\ [2mm]
\hline 
\hline
$pp \to S \to ZZ$ & $M_S = 250$ GeV & $M_S = 500$ GeV & $M_S = 750$ GeV & $M_S = 1$ TeV \\
\hline\hline
& & & & \\ [-3mm]
ATLAS (Run I)\cite{HH_ATL_ZZ_8tev} & 0.298 & 0.044  & 0.012 & 0.011  \\ [2mm]
\hline  
& & & & \\ [-3mm]
CMS (Run I)\cite{HH_CMS_WWandZZ_8tev} & 0.110  & 0.089 & 0.040 & 0.025 \\ [2mm]
\hline 
& & & & \\ [-3mm]
ATLAS (Run II)\cite{HH_ATL_ZZ_13tev} & 0.758 & 0.111  & 0.068 & 0.050  \\ [2mm]
\hline 
& & & & \\ [-3mm]
CMS (Run II)\cite{HH_CMS_ZZ_13tev} & 0.416 & 0.136  & 0.070 & 0.060  \\ [2mm]
\hline 
\end{tabular}
\caption{\footnotesize LHC 95\% upper limits on the cross-section, in pb, 
for a heavy scalar $S$ decaying to a $WW$ or a $ZZ$ pair, for the benchmark
values $M_S = 250$,
500, 750 and 1000 GeV respectively. In our work, we have used only the 
Run-2 data for the constraints. } 
\label{tab:HH}
\end{center}
\end{table}

\begin{figure}[!h]
\centerline{\includegraphics[width=0.9\textwidth]{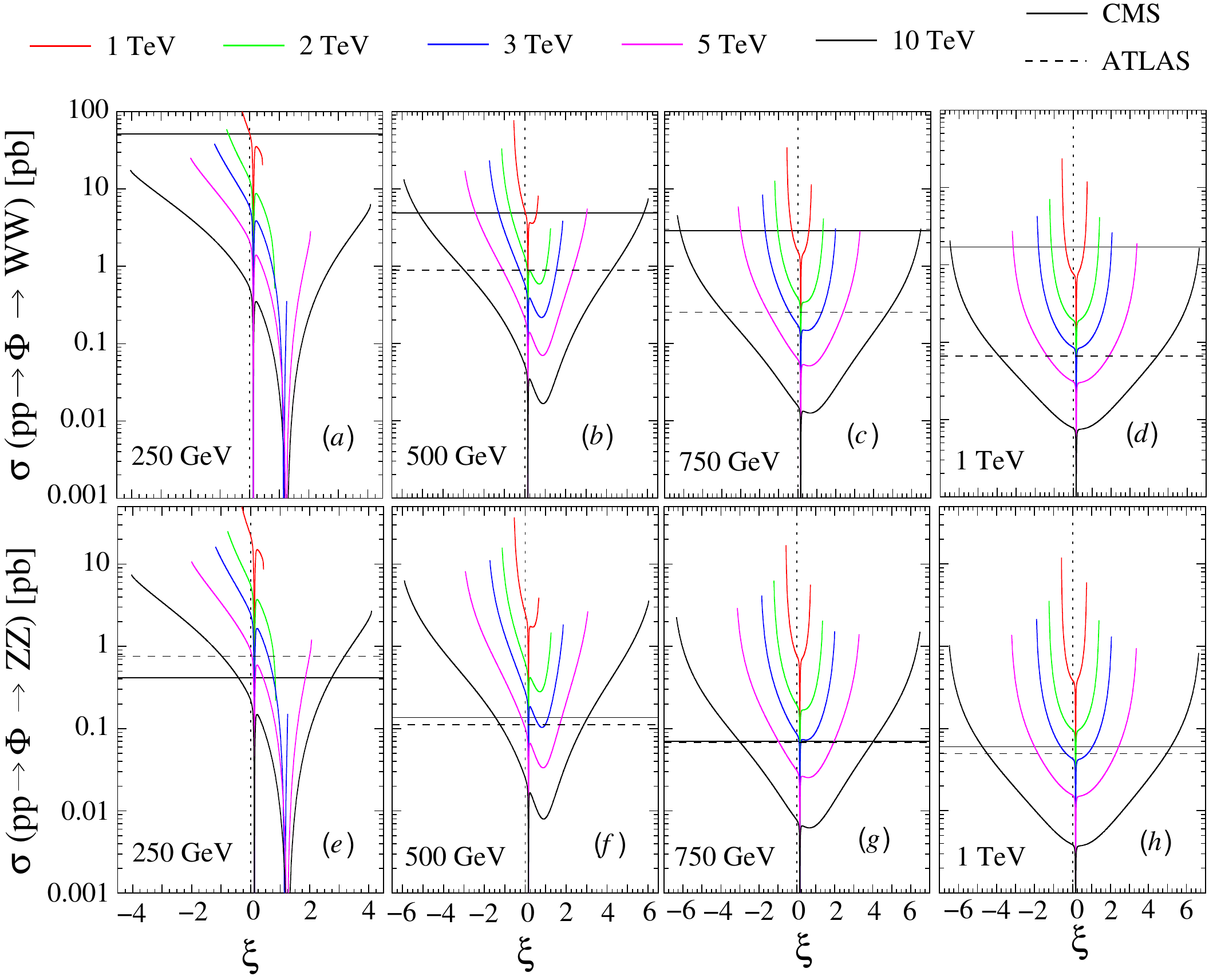} }
\vspace*{-0.2in}
\caption{\setstretch{1.05}\footnotesize Predictions of this model
vis-\'a-vis LHC searches for a heavy `SM-like' scalar. The upper set of
panels are for a $WW$ final state and the lower set of panels are for 
a $ZZ$ final state. Each panel shows the variation with $\xi$ for a
definite $M_\Phi$ as marked, and the different curves correspond to
different values of $\Lambda_\varphi$, as indicated in the legend above
the panels. Horizontal solid (dashed) lines indicate the 95\% C.L. 
CMS (ATLAS) 13~TeV constraints as in Table~\ref{tab:HH}.}
\label{fig:heavyHiggs}
\end{figure}

We are now in a position to compare these data with the predictions of our
theory. As in the case of the $H$ state, the cross section for 
$pp \to \Phi \to VV$, where $V = W,Z$, can be written
\begin{equation}
\sigma(pp \to \Phi \to VV) = \sigma(pp \to gg \to \Phi) 
\ {\mathcal B}(\Phi \to VV)
\label{eqn:HH2VV}
\end{equation}
where ${\mathcal B}(\Phi \to VV)$ is the branching ratio of the $\Phi$ to a 
$VV$ pair. These can be calculated in terms of the free parameters 
$\xi$, $M_\Phi$ and $\Lambda_\varphi$ respectively. Our results are shown
in Fig.~\ref{fig:heavyHiggs}. 

The four upper panels of Fig.~\ref{fig:heavyHiggs} represent the cross-section, in pb, for the process
$pp \to \Phi \to WW$ and the lower four panels represent the process
$pp \to \Phi \to ZZ$. In each row the panels correspond, from left to 
right, to $M_\Phi = 250$~GeV, 500~GeV, 750~GeV and 1~TeV, respectively.
Within each panel, the curves show the variation of the cross-section
with the mixing parameter $\xi$, for different values of the radion 
vacuum
expectation value, corresponding to different colours, as marked in the legend above
the panels. The horizontal solid lines correspond to the CMS bounds from
the 13~TeV data, as shown in Table.~\ref{tab:HH}, while the broken lines
correspond to the ATLAS 13~TeV data. 

All the curves have a distinct 
minimum at a small value of $\xi$ varying from 0.2 to 2 --- this corresponds
to a minimum in the cross-section $\sigma(pp \to gg \to \Phi)$ where
there is maximal cancellation in the amplitude for $gg \to \Phi$
due to the top quark loop and the trace anomaly term. In this region, the heavy scalar can be produced in association with a $W^\pm/Z$ and it further decays to $WW$ or $ZZ$ pairs,
leading to a final state with three gauge bosons or their decay products. 
In view of the low production cross-sections for higher values of $\Lambda_\varphi$, one has to consider hadronic decays of one or more of these gauge
bosons, and this immediately invites a large QCD background at the LHC. 
However, the region can be successfully probed at a high energy 
$e^+ e^-$ collider (such as the proposed ILC) with $\sqrt{s} = 1$~TeV \cite{Frank:2016oqi}.

In addition to the dip described above, there is a very sharp minimum,
very close to the vertical axis, which corresponds to the so-called 
`conformal' point, where $c_\Phi \to 0$. We defer the discussion of 
this point to the next section and focus here on the constraints obtainable
from the rest of the parameter space. Here, as in the case of signal 
strengths the constraints rule out larger values of $\xi$, with the exact
bound depending on the other two parameters of the theory.  

From Figs.~\ref{fig:signalstrengths} and \ref{fig:heavyHiggs} we can
draw some general conclusions. The first is that the effect of increasing
the mixing parameter $\xi$ becomes weaker and weaker as the vacuum
expectation value $\Lambda_\varphi$ keeps increasing. This is true both for the signal strengths in 
Fig.~\ref{fig:signalstrengths} as well the cross-section in Fig.~\ref{fig:heavyHiggs} and is easy to track down as due to the limiting
case $\gamma \to 0$. A similar argument may be made for the parameter
$M_\Phi$ -- at least numerically -- though the parameter dependence here is
much more complicated.
\begin{figure}[!h]
\centerline{\includegraphics[width=0.9\textwidth]{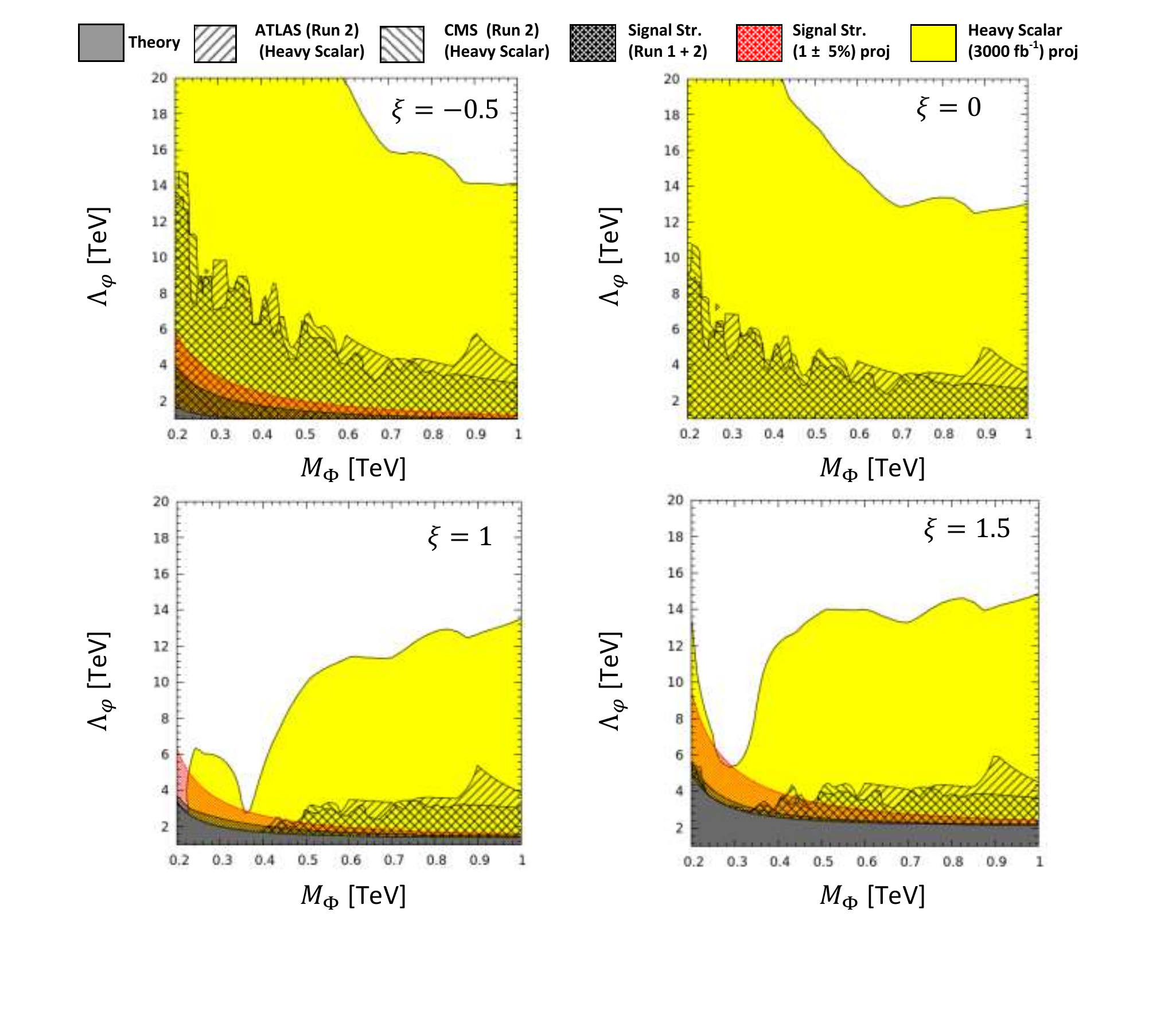} }
\vspace*{-0.6in}
\caption{\setstretch{1.05}\footnotesize Constraints from LHC data on the 
$\Lambda_\varphi$-$M_\Phi$ plane for different values of the mixing parameter
$\xi$. The region shaded grey is theoretically disallowed and the region
shaded dark grey is ruled out by the Higgs boson signal strengths. Hatching 
with opposite slants correspond to the ATLAS and CMS constraints from the
heavy scalar search. The red-shaded region represents a projection
of constraints from the signal strengths, assuming $\mu_{XX} = 1 \pm 0.05$ for all channels. Finally, the 
yellow-shaded region represents a combination of the ATLAS and CMS projected
discovery limits from the $ZZ$ channel, 
assuming a data collection of 3000~fb$^{-1}$ at 14~TeV.}
\label{fig:VEV-Mass}
\end{figure}
 We may argue, therefore, that for a fixed $\xi$,
the region with small $M_\Phi$ and small $\Lambda_\varphi$ is more constrained
--- which also corresponds to the commonsense argument that if these 
parameters are small, radion-mediated processes are large and vice versa.
These expectations are corroborated by our results shown in 
Fig.~\ref{fig:VEV-Mass}. Here we show the $\Lambda_\varphi$--$M_\Phi$ plane
for four different values of $\xi$, viz. $\xi = -0.5$, 0, 1 and 1.5, 
as marked
on each panel. As indicated in the key at the top, the region shaded grey
corresponds to the theoretically disallowed region, and includes all values of 
$\Lambda_\varphi < 1$~TeV, except in the panel on the top left, 
marked $\xi = 0$, which corresponds to the case of an un-mixed radion
of mass $M_\Phi$.  
Here, though values of $\Lambda_\varphi < 1$
TeV are theoretically permitted, the experimental constraints do not allow them, 
as is apparent from the figure. In all the panels, the dark grey shaded region
is ruled out by the signal strengths at Runs 1 and 2 and the hatched regions by the
ATLAS and CMS searches for a heavy scalar at Run-2 of the LHC. 
These are the strongest 
constraints and represent the state of the art as far as current 
experimental data are concerned\footnote{We have, in fact, considered
constraints from all the channels separately, but the others are
subsumed in the ones shown in the figure, and hence are not shown in
order to have uncluttered figures.}. The jagged shape of the curves reflects
the fact that the LHC has, till now, collected quite a small amount 
of data for rare processes like the decay of a heavy scalar. 
However, the LHC has the potential to
search much further, and this is shown by the red and yellow-shaded regions,
which represent, respectively, the expectations from the signal strength measurements if $\mu_{XX} = 1 \pm 0.05$ for all $X$, and the ATLAS and
CMS discovery limits at 95\% C.L. for the heavy $\Phi$ if the
LHC were to run at 14~TeV and collect 3000~fb$^{-1}$ of data\cite{HH_ZZ_ATL_14tev,HH_ZZ_CMS_14tev} --- which
may not be too far from the reality. For the panel with $\xi = 0$, there
are no constraints from the signal strengths, since the $H$ is completely
SM-like; but the constraints from the heavy scalar searches are quite strong
because that scalar is a pure radion. A comparative study of the four plots
indicates that the value $\xi \approx 1$ would permit the largest part of
the parameter space to survive consistently negative results from LHC, while
negative values of $\xi$ are better suited to a discovery of the heavy scalar
predicted in this theory.     

\begin{figure}[!ht]
\centerline{\includegraphics[width=0.9\textwidth]{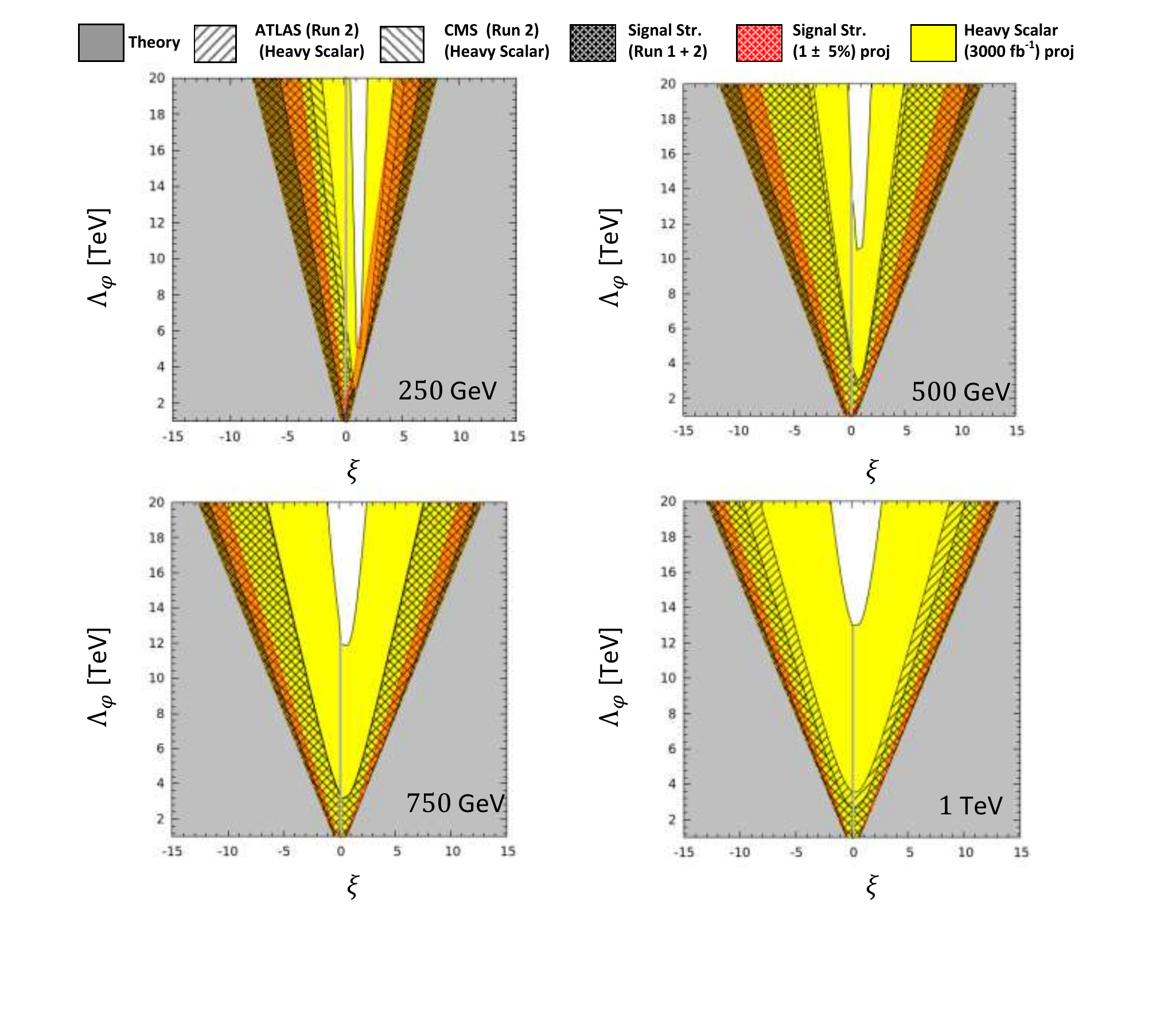} }
\vspace*{-0.7in}
\caption{\setstretch{1.05}\footnotesize Constraints from LHC data on the 
$\Lambda_\varphi$-$\xi$ plane for different values of the heavy scalar mass
$M_\Phi$. The region shaded grey is theoretically disallowed and the region
shaded dark grey is ruled out by the Higgs boson signal strengths. Hatching 
with opposite slants correspond to the ATLAS and CMS constraints from the
heavy scalar search. As in Fig.~\ref{fig:VEV-Mass}, the red-shaded region represents a projection
of constraints from the signal strengths, assuming $\mu_{XX} = 1 \pm 0.05$ for all channels and the yellow-shaded region represents a combination of the ATLAS and CMS projected
discovery limits, assuming a data collection of 3000~fb$^{-1}$ at 14~TeV.}
\label{fig:VEV-mix}
\end{figure}

Coming to constraints on $\xi$, it is clear from Figs.~\ref{fig:signalstrengths} 
and \ref{fig:heavyHiggs} that $\xi = 0$, which 
corresponds to the 125~GeV scalar being the Standard Model Higgs boson ---
not surprisingly --- is always allowed by the signal strength data. For given 
values of $M_\Phi$ and $\Lambda_\varphi$, 
$\xi$ can range on the positive and negative side, but when its magnitude
grows larger, all new physics effects grow and, at some point, higher
magnitudes of $\xi$ get disallowed -- first by the experimental constraints
and then by the requirement of theoretical consistency. For low values of 
$\Lambda_\varphi$ and $M_\Phi$, we arrive at this point for fairly low values of
$\xi$. As both these parameters increase, however, the allowed range
grows, creating a funnel-like shape, which grows wider as $\Lambda_\varphi$ and 
$M_\Phi$ increase. This is illustrated in Fig.~\ref{fig:VEV-mix}, where we 
show the $\Lambda_\varphi$-$\xi$ plane for the same choices of $M_\Phi$ as in
the earlier figures. The shading and hatching conventions of this figure
are exactly the same as those of Fig.~\ref{fig:VEV-Mass}. It is immediately
obvious that for low values of $\Lambda_\varphi$ close to 1~TeV, the range of
$\xi$ is severely constrained by theoretical consistency alone. A heavy 
scalar of mass 250~GeV is also rather severely constrained, except for 
a narrow cone, which will shrink further when the LHC finishes its run.
Constraints ease up for a heavier scalar, since that is much more difficult
to find. It is interesting that even if LHC completes its run without
finding any evidence for a heavy scalar up to 1~TeV, there will be a range
of parameter space where this model is still allowed. However, for these
parameters, the 125~GeV will be so similar to the SM Higgs boson, and the
interactions of the heavy scalar will be so heavily suppressed that the
model may no longer be interesting, at least from a phenomenological point 
of view.

An interesting feature of all the plots in Fig.~\ref{fig:VEV-mix} is the
needle-thin sliver of allowed parameter space which appears in every graph
close to the vertical axis. This corresponds, in every case, to the 
`conformal point' mentioned above, where all constraints from a heavy
scalar search weaken considerably. This region -- though extremely fine-tuned -- 
is interesting in its own right, and therefore we carry out a detailed study in
the next section. 

\bigskip

\noindent{\large\sc 4. The Conformal Point}

\begin{figure}[!htb]
\centerline{\includegraphics[width=1.05\textwidth]{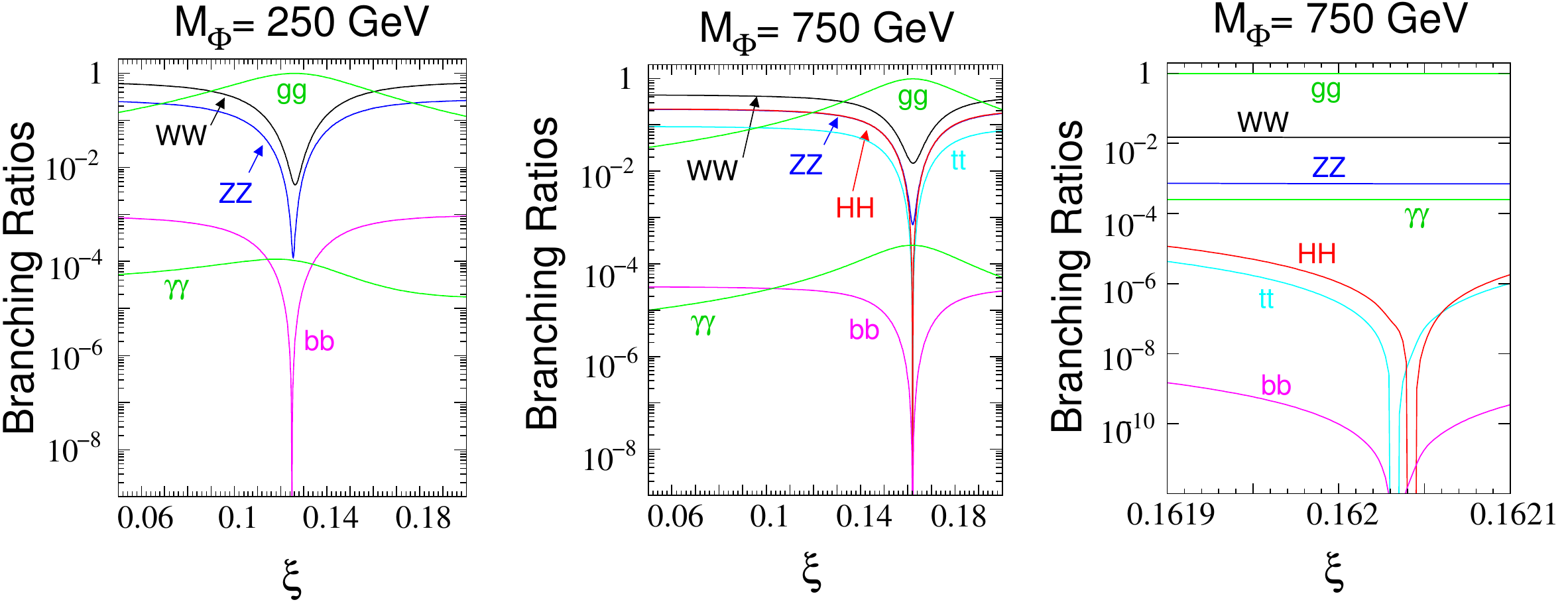} }
\vspace*{-0.1in}
\caption{\setstretch{1.05}\footnotesize Branching ratios of the heavy scalar
$\Phi$ in the neighbourhood of the conformal point. Note that the conformal
point is quite sensitive to the value of $M_\Phi$. There is some minor dependence 
on the radion vacuum
expectation value $\Lambda_\varphi$, but for purposes of comparison it has been
set at 2~TeV for every plot in this figure. The sharp drop in the tree-level
decays at the conformal point may be noted. The conformal point for the decay 
$\Phi \to HH$ is close to, but different from that for other decays, as is
clear in the panel on the right, which is a zoomed version of the central
panel.}
\label{fig:CnfPt}
\end{figure}

As explained before, for every choice of $M_\Phi$ and $\Lambda_\varphi$, there is 
a fixed value $\xi = \xi_0$ which satisfies the equation $c_\Phi = 0$, and 
hence
\begin{equation}
C(\xi) + \gamma A(\xi) = 0
\label{eqn:cnfpt}
\end{equation} 
and this is known as the `conformal' point\footnote{From this stage we drop
the quotes on `conformal'.}. It corresponds to the case when the tree-level
couplings $g_{\Phi X\bar{X}}$ of both the fermions and gauge bosons -- generically
denoted $X$ -- with the heavy scalar $\Phi$ vanish. This is a curious situation
and corresponds to the case when the mixing is fine-tuned to be such that the 
parts of the coupling arising from the SM $h$ and the radion $\varphi$ cancel 
each other. Like all fine-tuned situations, if this is the reality, it can
hardly
be a random effect, and must represent some deeper structure in the theory, which
is not addressed in our present formulation. Nevertheless, it is interesting to 
explore the phenomenological implications of this scenario. In this section,
therefore, we investigate the conformal point and see how it can be constrained using current and projected data, just as the other points can. 
It is important to note that though most of the tree-level couplings
of the $\Phi$ to pairs of SM particles vanish at the conformal point (except
for the coupling to $HH$ pairs), there exist one-loop couplings to pairs
of gauge bosons through the trace anomaly. This makes the pattern of 
branching ratios at the conformal point very different from that in
other regions of parameter space. The most important feature of this is
the fact that the decays $\Phi \to gg$ and $\Phi \to \gamma\gamma$ are 
considerably enhanced with respect to the others -- in fact the former
is the dominant decay mode. This behaviour is nicely exhibited in 
Fig.~\ref{fig:CnfPt}, where we exhibit the behaviour of the relevant 
branching ratios in the immediate vicinity of the conformal point. 
 
In Fig.~\ref{fig:CnfPt}, it is immediately apparent that for the particular
value $\xi = \xi_0$, the tree-level decay modes of $\Phi \to X\bar{X}$,
where $X$ is a massive gauge boson or a fermion, drop sharply by many orders of
magnitude. This is particularly true for the cases $X = t, b$ and $H$,
with the minimum for  the last case occurring at a slightly displaced
point from the others (best seen in the zoomed panel on the right). 
On the other hand, the branching ratios for the purely
one-loop decays, viz. $\Phi \to gg$ and $\Phi \to \gamma\gamma$ exhibit
a growth at the same point, attributable to their partial decay widths being
finite, whereas the others drop almost to zero. However, the decays to $WW$ and
$ZZ$ states do not disappear altogether because they too have anomaly
contributions. Naturally the decay $\Phi \to gg$ dominates the others because
of the appearance of the strong coupling as well as the colour factor. The
decay $\Phi \to \gamma\gamma$ also shows a gentle increase, but is intrinsically
much more rare than the digluon mode. At the conformal point, therefore, constraints on the model will have to be sought in a different 
fashion. One obvious way is to consider Higgs boson signal
strengths, for if the couplings of the $\Phi$ vanish that does not mean that 
the couplings of the $H$ will also vanish. 
Accordingly, there will be contributions to the signal
strengths and these can be used to constrain the model. In fact, even the
heavy scalar searches, i.e. $pp \to S \to VV$, where $V = W,Z$ can be used
to a limited extent, since the branching ratios $\Phi \to VV$, though small
at $\xi = \xi_0$, are not absolutely negligible. However -- and this is a
distinct feature of the conformal point -- the strongest bounds come from 
diphoton searches, which is not entirely surprising, given that this mode is 
considerably enhanced at the conformal point.  

\begin{figure}[!ht]
\begin{minipage}{0.55\textwidth}
\begin{center}
\includegraphics[width=\textwidth]{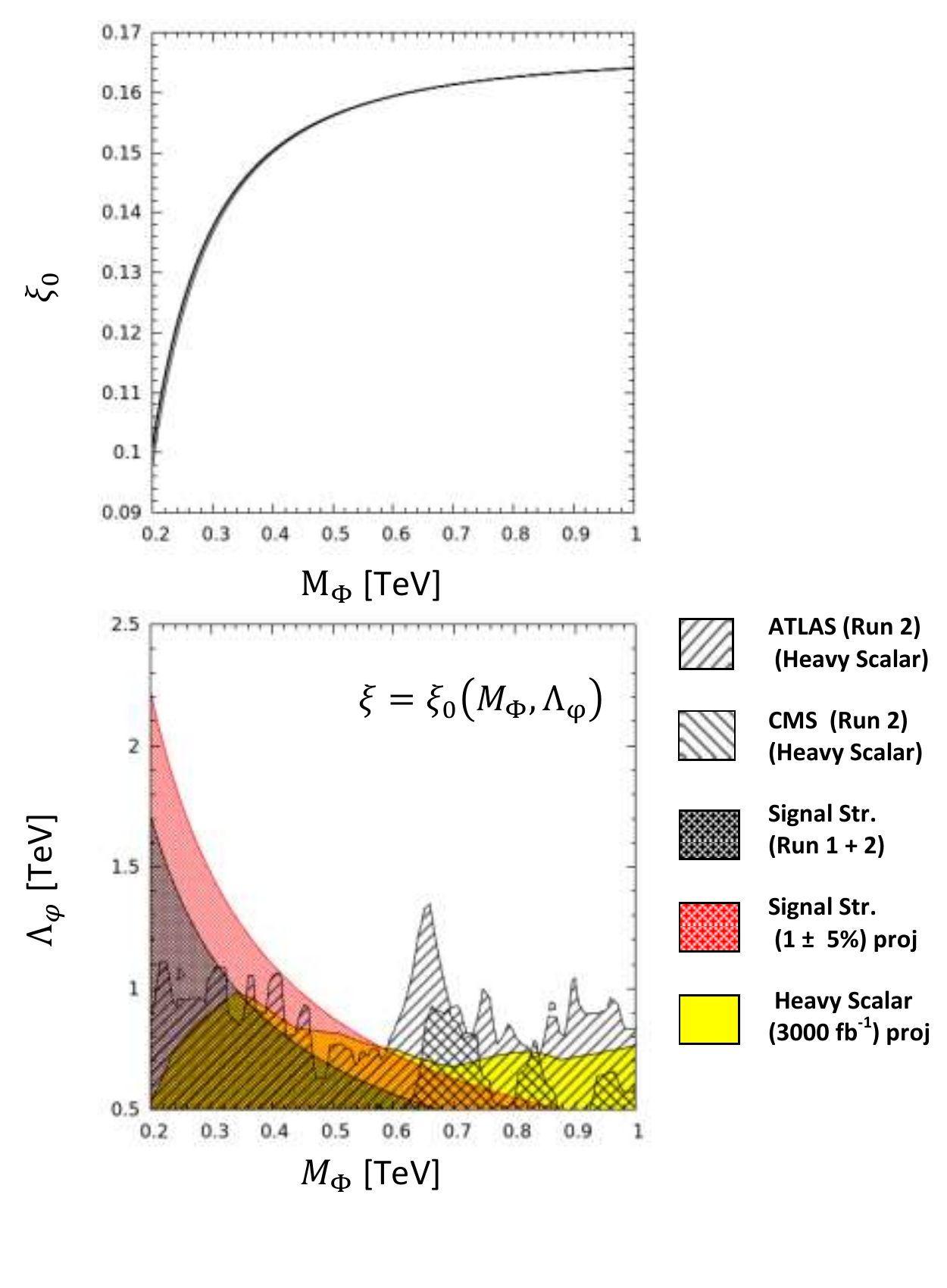}
\vspace*{-0.6in}
\caption{\setstretch{1.05}\footnotesize Constraints on the conformal point 
$\xi_0$. The variation of $\xi_0$ with $M_\Phi$ is shown in the upper panel. 
The thickness of the line corresponds to variation of $\Lambda_\varphi$ from
1 to 20~TeV. The lower panel shows the
$M_\Phi$--$\Lambda_\varphi$ plane, assuming that at every point the mixing parameter $\xi = \xi_0$. }
\label{fig:CnfBounds}
\end{center}
\end{minipage}
\hskip 0.01\textwidth
\begin{minipage}{0.44\textwidth}
\setstretch{1.15}
With these parameters, we now evaluate the measurables, viz. the signal strengths
and the cross-sections for $pp \to S \to VV$, where $V = W,Z$. These are then
compared with existing data to yield the constraints on the plane, as shown in
the lower panel of Fig.~\ref{fig:CnfBounds}. The conventions of this panel
are exactly the same as those of Figs.~\ref{fig:VEV-Mass} and \ref{fig:VEV-mix},
but the constraints follow a different pattern. As usual, low values of $M_\Phi$
and $\Lambda_\varphi$ are excluded. However, there are no theoretical constraints,
showing that there will always be a conformal point for any choice of model
parameters. For small values of $M_\Phi$, the strongest constraints come from
the signal strengths (dark grey shaded area), while for higher values, it is
the ATLAS and CMS data on diphotons -- not $WW$ and $ZZ$ -- from a heavy scalar
resonance, which yield the best constraints. Projecting signal strength 
measurements at the level of $\mu_{XX} = 1 \pm 0.05$ for all $X$ provides the
red-shaded band, showing that moderate 
\end{minipage}
\end{figure}

In trying to understand how the conformal point is constrained by the data,
we need to recognise that the conformal point $\xi_0$ is not unique, but a 
function of $M_\Phi$ and $\Lambda_\varphi$, with the dependence on the
former being much stronger than that on the latter. Its variation with $M_\Phi$ is shown
in the upper panel of  Fig.~\ref{fig:CnfBounds},  where the thickness
of the line corresponds to variation of $\Lambda_\varphi$
from 1~TeV to 20~TeV.  
This plot shows that the variation flattens out as $M_\Phi$
grows above 500~GeV, and has a very weak dependence on $\Lambda_\varphi$. Nevertheless, we have scanned a sizeable portion
of the $M_\Phi$--$\Lambda_\varphi$ plane
and calculated the values of $\xi_0$ at every point by solving 
Eq.~(\ref{eqn:cnfpt}).

\vspace*{-0.2in}
\vspace*{-0.2in}
improvement can be obtained if these 
measurements yield results much closer to the SM prediction.
The shaded yellow region represents the predictions from $ZZ$ decay modes of
a heavy scalar for the LHC running at 14~TeV with 3000~fb$^{-1}$\cite{HH_ZZ_ATL_14tev,HH_ZZ_CMS_14tev} of data  
(which is all that is currently available), and it does worse than the Run-2 data.
It may be expected that diphoton searches would provide better discovery limits --- when the Run-2 projections become available. 

All in all, we can conclude that the 
conformal point is somewhat less constrained than the rest of the parameter
space. It was this narrow window which had been used \cite{radion_conf} to 
explain the purported discovery of a heavy 750~GeV scalar during 2015-2016 
\cite{750GeV},
though that proto-signal did not survive the test of time\cite{{HH_ATL_gmgm_13tev,HH_CMS_gmgm_13tev}}. 

\bigskip
\noindent{\large\sc 5. Summary and Outlook}

The minimal Randall-Sundrum model continues to be one of the most elegant ways of
solving the hierarchy problem, and it works best if there is a Goldberger-Wise
stabilisation, which works best if there is a light radion state. Though
there are strong constraints on such a light radion per se, there remains room 
for a light radion mixed with the SM Higgs boson to survive. In this article, 
we have explored this possibility, using an existing formalism, in the light of 
current data from the LHC Runs 1 and 2. Our findings are summarised below.

The possibility of a radion-Higgs mixing arises essentially because we have no
independent measurement of the Higgs boson self coupling $\lambda$, so
that the SM formula $M_h^2 = 2\lambda v^2$ is open to other interpretations.
One of these is the mixed radion-Higgs scenario, where the lighter eigenstate 
is identified with the 125~GeV scalar discovered at the LHC. 
In this model, there are three free parameters,
viz. the mixing parameter $\xi$, the mass $M_\Phi$ of the heavy scalar $\Phi$,
and the radion vacuum expectation value $\Lambda_\varphi$. However, 
self-consistency of the theory imposes fairly stringent constraints on the 
choices of the mixing parameter $\xi$. These, as we show,
are further constrained by ($a$) the signal strengths measured for the decays
of the 125~GeV scalar at the LHC, and ($b$) the search for a heavy scalar
decaying into a pair of electroweak vector bosons, be they $W$'s, $Z$'s or 
photons. These lead to further bounds on the parameter space, essentially
pushing $\Lambda_\varphi$ above a TeV (and hence reducing all radion-mediated
effects) and $M_\Phi$ to values closer to a TeV, though here some avenues
for a lighter $M_\Phi$ remain. 

In addition to the current data, we have tried to predict discovery limits
at the LHC in two ways. One way is to use the signal strengths, and assume
that they will eventually converge within 5\% of the SM prediction. This
leads to modestly enhanced bounds on the radion-Higgs mixing scenario. The
other way is to use the projected discovery limits from the ATLAS and CMS 
Collaborations for a heavy scalar in Run-2, where we identify that heavy
scalar with our heavier eigenstate $\Phi$. This, in fact, is very effective
for most choices of the mixing parameter $\xi$ and is sensitive to rather high 
values of $M_\Phi$ and $\Lambda_\varphi$. The only exception is at the so-called
conformal point, which is a peculiar feature of this model, involving a value of 
the mixing parameter where the heavy scalar essentially decouples from SM fields.
Even this is constrained, however, by the signal strengths and by the diphoton
decay mode, which, being generated by the trace anomaly, survives the vanishing
of tree-level couplings. However, the smallest values of $M_\Phi$ and 
$\Lambda_\varphi$ are, indeed, allowed if this scenario were to be true.  

It is interesting to ask how our results would be modified if we replace
the simplistic model used above with a more phenomenologically-relevant
model where the fields can access the bulk. As explained in the Introduction,
the radion and Higgs fields, being still close to the TeV brane, mix in the same manner~\cite{bulk_radion}. The decay of the radion to the 
light quarks is severely suppressed because of the small overlap~\cite{Toharia:2008tm} of their wavefunctions in the bulk. Decays of the radion to massive gauge bosons are governed by an additional coupling that can be safely neglected for $\Lambda_{\varphi} \gsim 1$~TeV.  
Radions decaying to massless gauge boson pairs (especially to diphotons) 
is significantly enhanced, however, due 
to the tree-level coupling in the case of bulk scenario. However, this doesn't really effect our region of interest~\cite{Frank:2016oqi}. We
feel, therefore, that the results of this work are robust against
more realistic variations of the minimal model and may be safely adopted 
in such cases.   

To conclude, then, we have shown that a mixed radion-Higgs scenario is quite
consistent with the current experimental data at the LHC, and there is every
possibility that the heavy scalar predicted in this model could be discovered
as the LHC continues to run at its present energy of 13~TeV. Discovery of this
would certainly be one of the most exciting things to happen in the near future,
and, if, the branching ratios turn out to be consistent with this model, could
provide a powerful insight into the nature of spacetime itself. Such a happy
consummation is to be devoutly hoped for, but, for the present, we must reconcile
ourself to a fairly long wait as the Run-2 of the LHC continues. 

{\small
{\sl Acknowledgements}: The authors acknowledge useful discussions with Debjyoti 
Bardhan, Disha Bhatia and Abhishek Iyer. The work of SR was partly funded by the 
Board of Research in Nuclear Sciences, Government of India, under project no. 
2013/37C/37/BRNS.}
  

\end{document}